\documentclass[pra,aps,showpacs,floatfix,twocolumn]{revtex4-1}

\usepackage{amsmath}
\usepackage{amssymb}
\usepackage{graphicx}
\usepackage{dcolumn}
\usepackage{bm}
\usepackage[caption=false]{subfig}
\usepackage{multirow}
\usepackage{xcolor}
\usepackage{latexsym}
\usepackage[normalem]{ulem} 
\usepackage{wrapfig}

\begin{document}

\title{Quantum-classical correspondence in the vicinity of periodic orbits}

\author{Meenu Kumari$^{1,2}$ and Shohini Ghose$^{1,3,4}$}
\affiliation{$^1$Institute for Quantum Computing, University of Waterloo, Canada N2L 3G1} 
\affiliation{$^2$Department of Physics and Astronomy, University of Waterloo, Canada N2L 3G1}
\affiliation{$^3$Department of Physics and Computer Science, Wilfrid Laurier University, Waterloo,
Canada N2L 3C5} 
\affiliation{$^4$Perimeter Institute for Theoretical Physics, 31 Caroline St N, Waterloo, Ontario, N2L 2Y5, Canada}

\date{\today}

\begin{abstract}
Quantum-classical correspondence in chaotic systems is a long-standing problem. We describe a method to quantify Bohr's correspondence principle and calculate the size of quantum numbers for which we can expect to observe quantum-classical correspondence near periodic orbits of Floquet systems. Our method shows how the stability of classical periodic orbits affects quantum dynamics. We demonstrate our method by analyzing quantum-classical correspondence in the quantum kicked top (QKT), which exhibits both regular and chaotic behavior.  We use our correspondence conditions to identify signatures of classical bifurcations even in a deep quantum regime. Our method can be used to explain the breakdown of quantum-classical correspondence in chaotic systems.
\end{abstract}


\maketitle

\section{\label{sec:level1}Introduction}
Quantum-classical correspondence and the quantum-to-classical transition have been topics of fundamental interest since the birth of quantum theory in the early 20th Century. The connection between quantum and classical mechanics remains a partially understood subject, particularly in chaotic systems. According to the correspondence principle, the predictions of quantum physics should agree with the predictions of classical physics in appropriate limits wherever classical physics is applicable. There are multiple ways in which the correspondence principle has been formulated. These include Bohr's correspondence principle \cite{nielsen2013correspondence}, Ehrenfest's theorem \cite{Ehrenfest1927}, and Liouville correspondence \cite{Joshua1997a,Joshua1997b}, with each having its own subtleties \cite{Ballentine1994,gao1999breakdown,makowski2002bohr}.  According to Bohr, quantum-classical correspondence is attained in the limit of large quantum numbers (or when $\hbar \rightarrow 0$ relative to the phase space of the dynamics). According to Ehrenfest's theorem, the evolution of expectation values of observables in a quantum system should coincide with the corresponding classical evolution until a time known as Ehrenfest's time ($t_{\text{EH}}$) that depends on the system dynamics. While $t_{\text{EH}}$ is  large for regular systems in the semiclassical limit, it can be very small for chaotic systems even in the semiclassical limit \cite{berman1978,Zurek1994,Zurek1995}.  

Since Bohr's correspondence principle involves large quantum numbers, a natural question that arises is how large is large enough to see correspondence. Another important question is why correspondence breaks down in chaotic systems. In this paper, we address both these questions by analyzing the effect of stability and bifurcations of classical periodic orbits on quantum dynamics. We explore quantum-classical correspondence in the quantum kicked top (QKT) - a multiqubit time-periodic system that is a standard paradigm for exploring chaos \cite{haake1987}. This system is of particular interest because it displays bifurcations, regular behavior as well as chaotic behavior in the classical limit, and is one of the few chaotic systems that has been experimentally realized in the quantum regime \cite{chaudhury2009nature,neill2016ergodic}. Furthermore, since it is finite-dimensional, there are no truncation errors in the study of this system. 

Our study is based on an analysis of classical periodic orbits and their stability. We thus first present a classical periodic orbit analysis and bifurcation study of the kicked top. We then provide criteria for calculating the quantum number (in this case, the collective qubit spin $j$) for which the quantum dynamics of a state localized on a periodic orbit will correspond to the classical dynamics. Our criteria are based on the orthogonality of quantum states centered on the different points of the periodic orbits. When the criteria are satisfied, signatures of the classical bifurcations are clearly reflected in the quantum dynamics, even in a deep quantum regime. These signatures become more pronounced in a semiclassical regime. Furthermore, we show that in chaotic systems, the existence of orbits of very high periodicity can lead to a violation of our criteria and thus result in a short Ehrenfest break time. Studies of quantum-classical correspondence for systems with a mixed phase space of periodic islands and chaotic regions are more challenging compared to purely regular or chaotic systems. Our approach is thus particularly useful in the analysis of such mixed systems. 

The paper is organized as follows. In Sec. \ref{sec:level2}, we briefly describe the quantum kicked top model. Section \ref{sec:level3} includes a classical analysis of the kicked top with explicit calculations of periodic orbits and bifurcations. In Sec. \ref{sec:level4}, we describe our criteria for the quantification of Bohr's correspondence principle in periodic Floquet systems. In Sec. \ref{sec:level5}, we apply our criteria to the QKT. We show that when the criteria are satisfied, quantum-classical correspondence is evident even in a deep quantum regime. We also illustrate the effect of instability of classical periodic orbits on the quantum dynamics. In Sec. \ref{sec:level6}, we use our criteria to identify new quantum signatures of classical bifurcations in the kicked top dynamics, in a deep quantum regime as well as in the semiclassical regime. 
In Sec. \ref{sec:level7}, we discuss how our criteria can be used to explain the divergence of quantum and classical dynamics in chaotic systems. Finally, we present a summary of our results in Sec. \ref{sec:level8}.

\section{\label{sec:level2}Background}

\subsection{\label{sec:level2a}The quantum kicked top}
\begin{figure}[t] 
\centering{\includegraphics[scale=0.6]{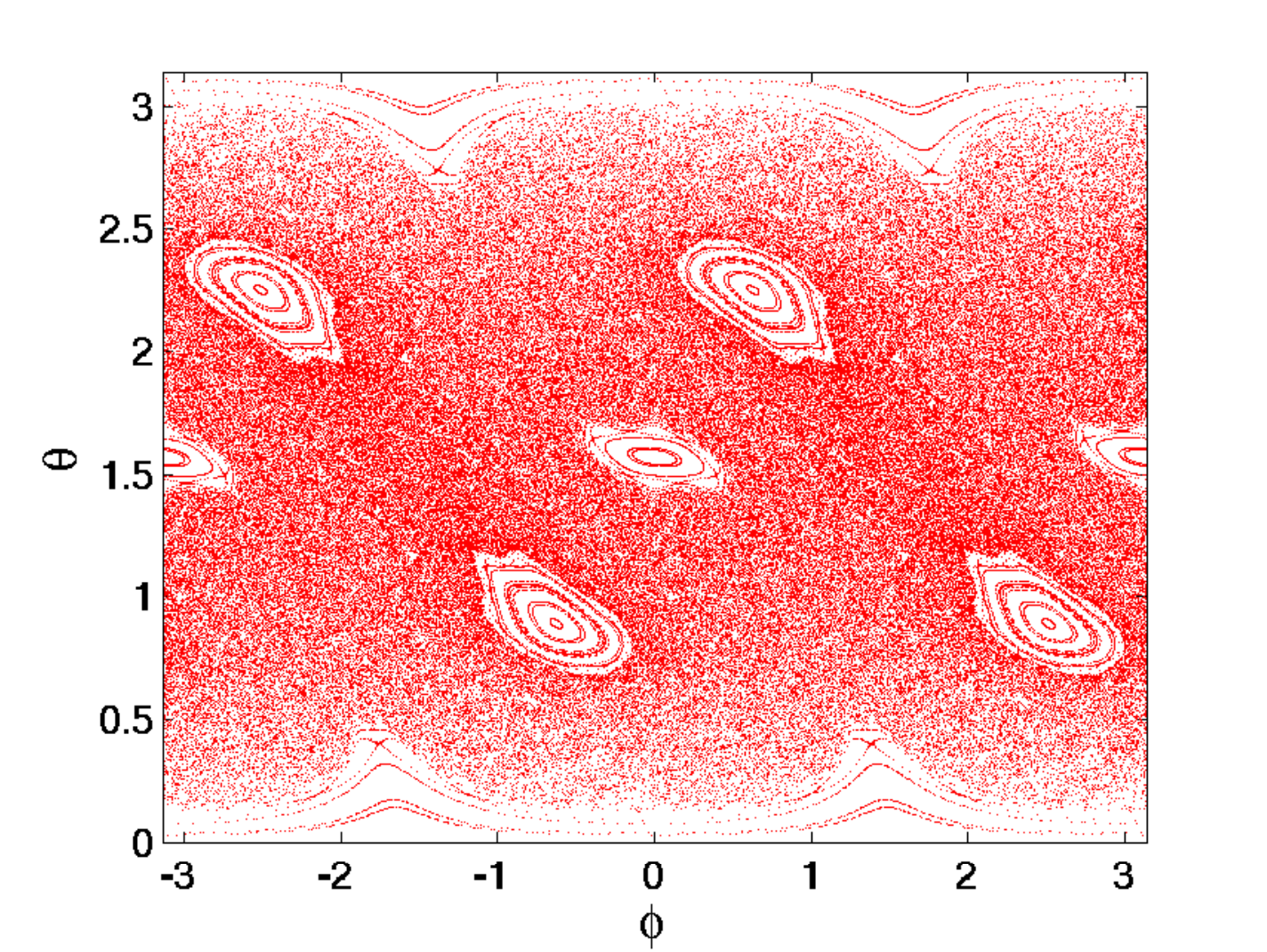}}
\caption{Classical stroboscopic phase space of the kicked top for parameter values $\kappa = 3.0$, $\tau = 1.0, p = \frac{\pi}{2}$. $\theta$ and $\phi$ are plotted after each kick for 1360 initial conditions, each evolved for 150 kicks.}
\label{phase_space}
\end{figure}

The quantum kicked top is a time-dependent periodic  system governed by the Hamiltonian \cite{haake1987}
\begin{equation}
H = \hbar \frac{\kappa}{2 j \tau} J_z^2 + \hbar p J_y \sum_{n= - \infty}^{\infty} \delta (t - n \tau),
\label{top1}
\end{equation}
where $J_x,J_y$ and $J_z$ are angular momentum operators. Since the square of the angular momentum operator commutes with the Hamiltonian ($[H,J^2] = 0$), its eigenvalue $j(j+1)\hbar^2$, and thus $j$, is a constant of motion. The Floquet time evolution operator for one time period, $\tau$, is
\begin{equation}
U=\exp{(-i \frac{\kappa}{2j \tau}J_z^2)} \exp{(-ipJ_y)}.
\label{top2}
\end{equation}
Each time period consists of a linear rotation by angle $p$ about the $y$ axis and a nonlinear rotation about the $z$ axis. The classical dynamics can be obtained by writing the Heisenberg equations of motion for the angular momentum operators and then taking the limit $j \rightarrow \infty$ \cite{haake1987}. Setting $X=J_x/j$, $Y=J_y/j$ and $Z=J_z/j$, the classical equations of motion for $p=\pi/2$ are
\begin{eqnarray}
X((n+1)\tau) & = & Z(n\tau) \cos{(\kappa X(n\tau))} + Y(n\tau) \sin{(\kappa X(n\tau))}, \nonumber \\
Y((n+1)\tau) & = & -Z(n\tau) \sin{(\kappa X(n\tau))} + Y(n\tau) \cos{(\kappa X(n\tau))}, \nonumber \\
Z((n+1)\tau) & = & -X(n\tau).
\end{eqnarray}
As the chaoticity parameter, $\kappa$, is varied from 0 to 7, the classical dynamics ranges from fully regular motion (for $\kappa \lesssim 2.1 $) to a mixture of regular and chaotic behavior for different initial conditions (for $ 2.1 \lesssim \kappa \lesssim 4.4$) to fully chaotic motion (for $\kappa \gtrsim 4.4$). The classical stroboscopic map (in polar co-ordinates) for a range of initial conditions with $\kappa=3$ is shown in Fig. \ref{phase_space}.

Quantum mechanically, we can view the quantum kicked top as a multiqubit system of $N=2j$ qubits. The symmetric subspace of the $2j$ qubits has constant angular momentum $j$, and the kicked top evolves in this subspace spanned by the eigenstates of $J^2$ and $J_z$, $|j,m\rangle, m=\{-j,-j-1,....j\}$.  

\subsection{\label{sec:level2b}Spin coherent states (SCS)}
Coherent states are minimum uncertainty states and are thus the closest quantum analog of classical states. For spin systems, these are the so-called spin coherent states (SCS) \cite{Arecchi1972}. Given any point $(\theta, \phi)$ in the classical phase space, we can construct SCS states $|\theta,\phi \rangle$ by applying the rotation operator $R(\theta,\phi) = \exp{[i \theta(J_x \sin{\phi} - J_y \cos{\phi})]}$ on the state $|j,j\rangle$, $|\theta, \phi \rangle = R(\theta, \phi)|j,j \rangle$. This yields a minimum uncertainty state centered on the point $(\theta, \phi)$, that is, the expectation value of the angular momentum of this state is $(j \sin{\theta}\cos{\phi}, j \sin{\theta}\sin{\phi}, j \cos{\theta})$. The uncertainty of this state is $(\langle \textbf{J}^2 \rangle - \langle \textbf{J} \rangle^2) / j^2 = 1/j$. Thus, for larger $j$ values, the SCS becomes highly localized at the point $(\theta, \phi)$ in phase space and better approximates the classical states. 

\subsection{\label{sec:level2c}Husimi phase space distribution}
We study the quantum evolution of the kicked top in phase space using the Husimi phase space distribution function \cite{HusimiDist1981}. Given any angular momentum quantum state $\rho$, the Husimi distribution is given by, 
\begin{equation}
Q(\theta,\phi) = \frac{2J+1}{4 \pi} \langle \theta, \phi |\rho | \theta, \phi \rangle,
\label{husimi1}
\end{equation}
which is equal to $\frac{2J+1}{4 \pi} |\langle \theta, \phi | \psi \rangle|^2$ for pure states.

\section{\label{sec:level3}Classical analysis of the kicked top: Periodic orbits and bifurcations}
We begin by analyzing the existence and stability of classical fixed points and some 2-periodic and 4-periodic orbits as the parameter $\kappa$ is varied with $p=\pi/2$.
To study the stability of a period-$n$ orbit, we calculate eigenvalues of the Jacobian of $F^n$ at the period-$n$ point, where $\left(X((n+1)\tau),Y((n+1)\tau),Z((n+1)\tau)\right) = F(X(n\tau),Y(n\tau),Z(n\tau))$. If $|\lambda| \leq 1$ $\forall \lambda$, then the period-$n$ orbit is stable, otherwise it is unstable. Table \ref{CP} lists a few interesting fixed points and periodic orbits of the kicked top.
The variable $x_0$ in $FP_1$, $FP_2$ and $P2_A$ in Table \ref{CP} is obtained from the normalization condition, 
\begin{equation}
2x_0^2 + \frac{[x_0 \sin{(\kappa x_0)}]^2}{[1-\cos{(\kappa x_0)]^2}} = 1.
\label{normalization}
\end{equation}

  \begin{table*}[t]
        \begin{tabular}{ccl}
        $FP_1$ & = & $(0,1,0)$  \\
        $FP_2$ & = & $(0,-1,0)$ \\
        $FP_3$ & = & $\left(x_0,\dfrac{x_0 \sin{(\kappa x_0)}}{(1-\cos{(\kappa x_0))}},-x_0 \right)$ \\
$FP_4$ & = & $\left(-x_0,\dfrac{x_0 \sin{(\kappa x_0)}}{(1-\cos{(\kappa x_0))}},x_0 \right)$ \\
$P2_A$ & = & $\left(x_0,-\dfrac{x_0 \sin{(\kappa x_0)}}{(1-\cos{(\kappa x_0))}},x_0 \right) \leftrightarrow \left(-x_0,-\dfrac{x_0 \sin{(\kappa x_0)}}{(1-\cos{(\kappa x_0))}},-x_0 \right)$ \\
$P4$ & = & $(1,0,0) \rightarrow (0,0,-1) \rightarrow (-1,0,0) \rightarrow (0,0,1) \rightarrow (1,0,0)$ \\
$P2_B$ & = & $\left(\dfrac{\pi}{\kappa},\sqrt{1-2(\dfrac{\pi}{\kappa})^2},\dfrac{\pi}{\kappa} \right) \leftrightarrow \left(-\dfrac{\pi}{\kappa},-\sqrt{1-2(\dfrac{\pi}{\kappa})^2},-\dfrac{\pi}{\kappa} \right)$ \\
$P2_C$ & = & $\left(-\dfrac{\pi}{\kappa},\sqrt{1-2(\dfrac{\pi}{\kappa})^2},\dfrac{\pi}{\kappa} \right) \leftrightarrow \left(-\dfrac{\pi}{\kappa},-\sqrt{1-2(\dfrac{\pi}{\kappa})^2},\dfrac{\pi}{\kappa} \right)$  \\
$P2_D$ & = & $\left(\dfrac{\pi}{\kappa},\sqrt{1-2(\dfrac{\pi}{\kappa})^2},-\dfrac{\pi}{\kappa} \right) \leftrightarrow \left(\dfrac{\pi}{\kappa},-\sqrt{1-2(\dfrac{\pi}{\kappa})^2},-\dfrac{\pi}{\kappa} \right)$ \\
$P2_E$ & = & $\left(\dfrac{\pi}{\kappa},-\sqrt{1-2(\dfrac{\pi}{\kappa})^2},\dfrac{\pi}{\kappa}\right) \leftrightarrow \left(-\dfrac{\pi}{\kappa},\sqrt{1-2(\dfrac{\pi}{\kappa})^2},-\dfrac{\pi}{\kappa} \right)$
    \end{tabular}
    \caption{Fixed points and periodic orbits of kicked top (in the form $(X,Y,Z)$)}
    \label{CP}

  \end{table*}

$FP_1$ and $FP_2$ are fixed points for all values of $\kappa$. The eigenvalues of the Jacobian at $(0,1,0)$ are $\left(1,\dfrac{\kappa+\sqrt{\kappa^2-4}}{2},\dfrac{\kappa-\sqrt{\kappa^2-4}}{2}\right)$. Clearly, for $\kappa > 2$, the eigenvalue, $\dfrac{\kappa+\sqrt{\kappa^2-4}}{2} > 1$. Thus, this fixed point loses stability at $\kappa=2$, which implies that $\kappa=2$ is a bifurcation point. 

\begin{figure}
    \centering
    \includegraphics[width=0.5\textwidth]{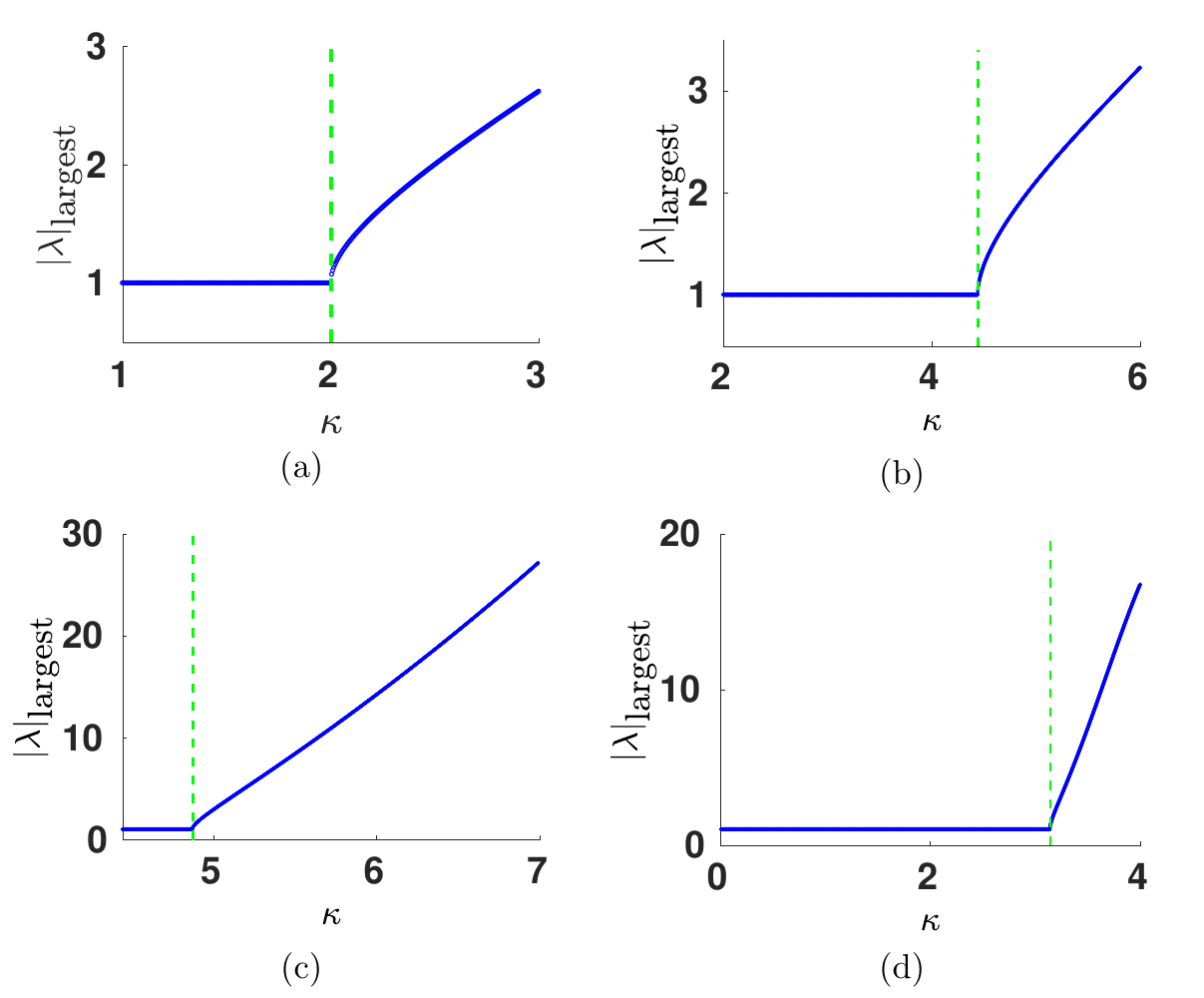}
    \caption{(a) Largest eigenvalue of the Jacobian matrix of $F$ at $FP_1$  as a function of $\kappa$ showing loss of stability of $FP_1$ at $\kappa=2$. (b) Largest eigenvalue of the Jacobian matrix of $F$ at $FP_3$ as a function of $\kappa$ showing loss of stability of $FP_3$ at $\kappa= \sqrt{2}\pi$. (c) Largest eigenvalue of the Jacobian matrix of $F^2$ at $P2_B$ as a function of $\kappa$ showing loss of stability of $P2_B$ at $\kappa \approx 4.8725$. (d) Largest eigenvalue of the Jacobian matrix of $F^4$ at $P4$ as a function of $\kappa$ showing loss of stability of $P4$ at $\kappa =\pi$.  Vertical dashed lines in the four plots represent the parameter value at which loss of stability occurs.}
    \label{bifurcation1}
\end{figure}

\begin{figure}
\centering\includegraphics[width=0.5\textwidth]{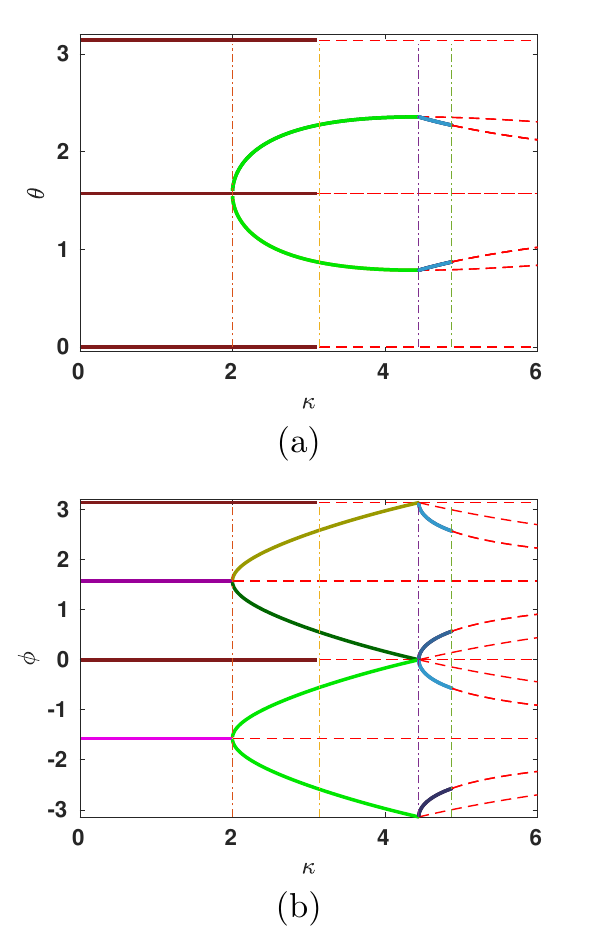}
\caption{Classical bifurcation diagram (a) $\theta$ vs $\kappa$, (b) $\phi$ vs $\kappa$. All solid lines represent stable fixed points and periodic orbits. Dashed lines represent unstable fixed points and periodic orbits. Dashed dotted vertical lines represent bifurcation parameter values.}
\label{bifur_diagram}
\end{figure}

At $\kappa=2$,  $FP_1$ gives rise to two fixed points: $FP_3$ and $FP_4$. $FP_2$ becomes a period-2 orbit, $P2_A$. $FP_3$, $FP_4$ and $P2_A$ lose stability at $\kappa= \sqrt{2}\pi$ (Fig.~\ref{bifurcation1}) and give rise to four stable period-2 orbits: $P2_B, P2_C, P2_D \mbox{ and } P2_E$. These four period-2 orbits lose stability at $\kappa \approx 4.8725$ (Fig.~\ref{bifurcation1}). There exists a period-4 orbit, $P4$, at all values of $\kappa$. It loses its stability at $\kappa=\pi$ (Fig. \ref{bifurcation1}). Figure ~\ref{bifur_diagram} shows the bifurcation diagram for the mentioned periodic orbits, explicitly showing the bifurcation points $\kappa =2, \pi, \sqrt{2}\pi, 4.8725$.

\section{\label{sec:level4}Quantifying Bohr's correspondence principle}
Bohr's correspondence principle broadly states that quantum dynamics will approach classical dynamics in the limit of large quantum numbers. Our goal is to quantify how large the quantum numbers need to be to observe similarity in classical and quantum dynamics. Here, we provide a quantification method based on periodic orbits for Floquet systems, which are periodically driven systems. We propose that the quantum dynamics in the vicinity of any classical period-$n$ orbit for such systems will be similar to the classical dynamics when:
\begin{enumerate}
\item the coherent states centered on all the $n$ points in the period-$n$ orbit are orthogonal to each other.
\item the coherent states centered on multiple periodic orbits that are related by the symmetries of the system are orthogonal to each other. 
\end{enumerate}

We note that the existence of symmetries in the system may lead to quantum mechanical phenomena between the periodic orbits related by these symmetries, such as, for example, dynamical tunneling \cite{davis1981quantum,keshavamurthy2011dynamical}. If so, then the conditions described above will not be sufficient to ensure correspondence in a deeply quantum regime. 

The two conjectured criteria above can be understood in the following way. In the limit where the classical states are distinguishable points in phase space, the classical dynamics evolves in a localized manner among these classical states in the phase space. At the quantum level, distinguishability is associated with orthogonality of quantum states. Consider the quantum dynamics of the same system with the initial state being a coherent state localized at one of the points in a stable period-$n$ orbit. We would expect the quantum dynamics to be similar to the classical dynamics if the quantum state evolves in a localized manner similar to the classical evolution. This localized evolution could occur if at any time in the evolution, the quantum state has high overlap with a coherent state centered at one of the classical points of the period-$n$ orbit and negligible overlap with coherent states centered on the rest of the points. This can be assured if the set of coherent states centered at the points of a period-$n$ orbit form an orthogonal set. However, if this set is a nonorthogonal set, then high survival probability at any classical point may still allow a significant amount of survival probability at other classical points in the period-$n$ orbit as well and thus may generally cause a departure from classical dynamics.

The richness of the classical phase space determines the quantum number (or the effective Planck's constant value) at which there will be a correspondence between classical and quantum dynamics.  For lower quantum numbers, the size of the Hilbert space restricts the dimension of any set that consists of states orthogonal to each other. Therefore, in systems whose classical phase space has only fixed points and period-$n$ orbits with small $n$, correspondence in the deep quantum regime is more likely to occur. We will illustrate our conjecture in the kicked top in the next section.

The overlap between any two spin coherent states, $| \theta, \phi \rangle$ and $| \theta_0, \phi_0 \rangle $,  is given by \cite{lombardi2011}:
\begin{equation}
|\langle \theta, \phi | \theta_0, \phi_0 \rangle| = \left(\cos{ \left[\frac{\chi(\theta \phi, \theta_0 \phi_0)}{2} \right]} \right)^{2J},
\label{overlap1}
\end{equation}
where $\chi(\theta \phi, \theta_0 \phi_0)$ is the angle between the direction vectors, $(\theta, \phi)$ and $(\theta_0, \phi_0)$ on the unit sphere, $\mathbb{S}^2$. Thus, Eq.~(\ref{overlap1}) is a handy tool to calculate the orthogonality of the spin coherent states for our quantification criteria.

\section{\label{sec:level5}Quantum versus classical dynamics of the kicked top}

\begin{figure*}
    \centering
    \includegraphics[scale=1.53]{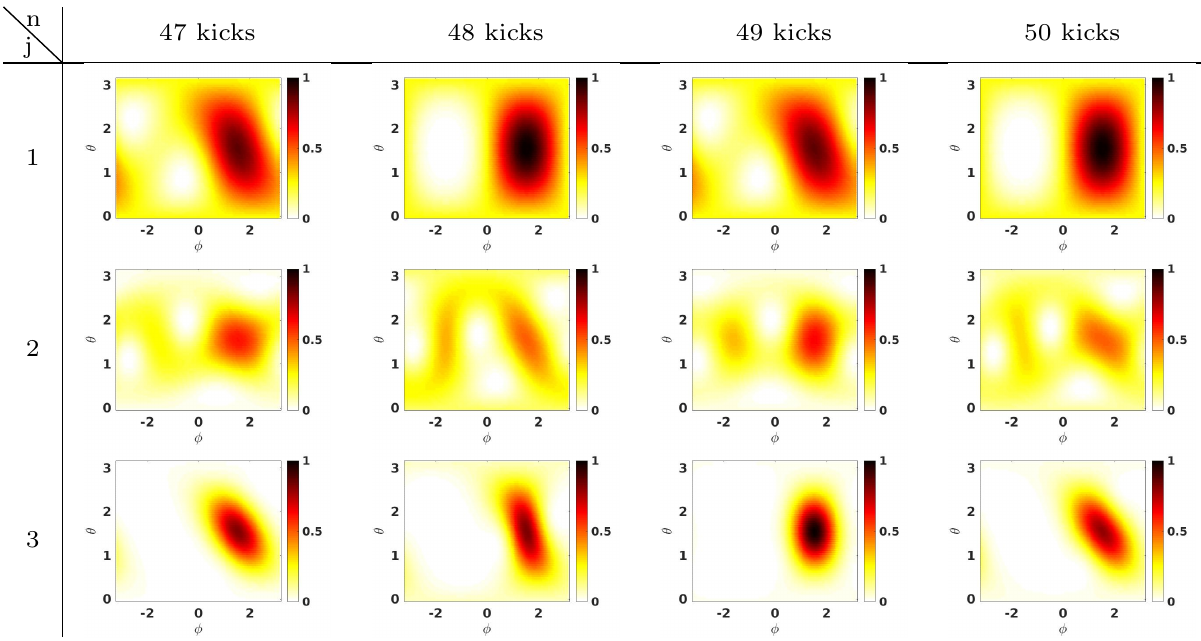}
    \caption{Evolution of the Husimi phase space distribution of an SCS centered on $FP_1$ [$(X,Y,Z)=(0,1,0)$] for three different $j$ values with $\kappa=1.5$. Like the classical dynamics, the quantum dynamics remains localized at $FP_1$, even in a deeply quantum regime, except for $j=2$ when dynamical tunneling to $FP_2$ occurs.} 
    \label{k1}
\end{figure*}

In sec. \ref{sec:level3}, we showed that in the QKT, the number of fixed points and periodic orbits increases as the chaoticity parameter, $\kappa$, is increased (keeping the parameter $p=\pi/2$ fixed), owing to the many bifurcations. In this section, we illustrate our quantification criteria in the kicked top.

\subsection{\label{sec:level5a}$\kappa<2$}

For $\kappa <2$, the only periodic orbits in the classical phase space are $FP_1$, $FP_2$ and $P4$. $FP_1$ and $FP_2$ are isolated fixed points with no other periodic orbit in their vicinity in the phase space. The spin coherent states centered on $FP_1$ and $FP_2$ are orthogonal to each other for all $j$ values. Thus, we observe correspondence between the classical and quantum dynamics at these fixed points, $FP_1$ and $FP_2$, even for a very low quantum number, $j=1$, as illustrated in Fig. \ref{k1}. However, in this deep quantum regime, there is the possibility of dynamical tunneling since
$FP_1$ and $FP_2$ are related by a symmetry of the square of the kicked top map for $p=\frac{\pi}{2}$, that is rotation by angle $\pi$ around the $x$-axis \cite{haake1987,Sanders1989}. This dynamical tunneling between the two fixed points, can be observed for some small values of $j$ (for example $j=2$ in Fig. \ref{k1}) but as the value of $j$ is increased further, the correspondence is recovered.

\begin{figure*}
    \centering
    \includegraphics[scale=1.53]{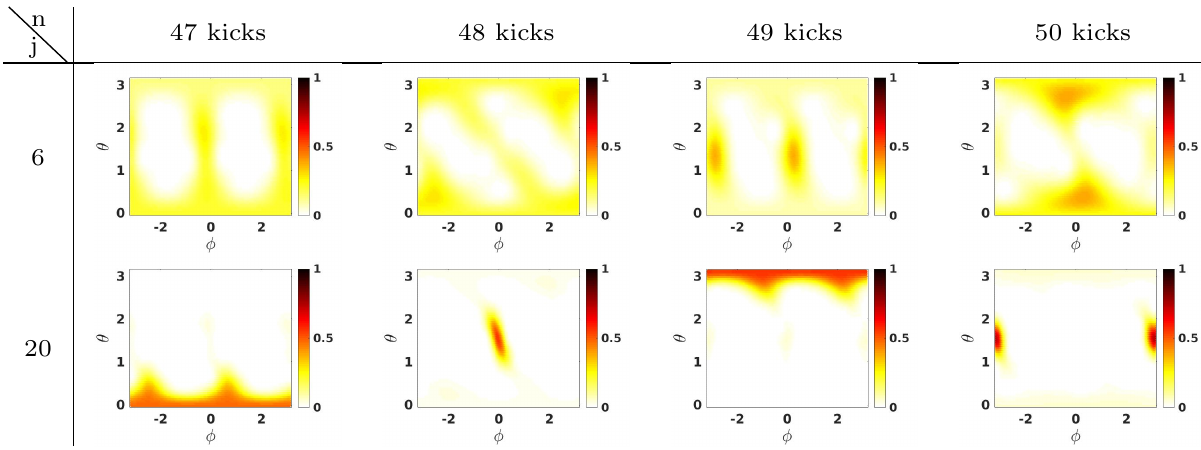}
    \caption{Evolution of the Husimi phase space distribution of an SCS centered on a point in $P4$ ($(X,Y,Z)=(1,0,0)$) for 2 different $j$ values with $\kappa=1.5$. For $j=6$ (first row), the quantum dynamics does not correspond to the classical dynamics, but for $j=20$, there is clear correspondence.} 
    \label{k2}
\end{figure*}

In contrast, the spin coherent states centered on the four points in the $P4$ orbit are not orthogonal to each other for very small $j$ values. From Eq.(\ref{overlap1}), the overlap between the spin coherent states at any two consecutive points in this period-4 orbit is given by $\left(\frac{1}{\sqrt{2}} \right)^{2j}$, which is of the order $10^{-7}$ for $j=20$, and $\approx 0.156$ for $j=6$. Thus, for $j$ values $\lesssim 20$, we do not see quantum-classical correspondence if we start at any one of the period-4 points, but we do see correspondence for large enough $j$ values $\gtrsim 20$, as illustrated in Fig. \ref{k2}.

\subsection{\label{sec:level5b}$2 < \kappa< \pi$}
\def \scaleval {.21}

\begin{figure*}
    \centering
    \includegraphics[scale=1.53]{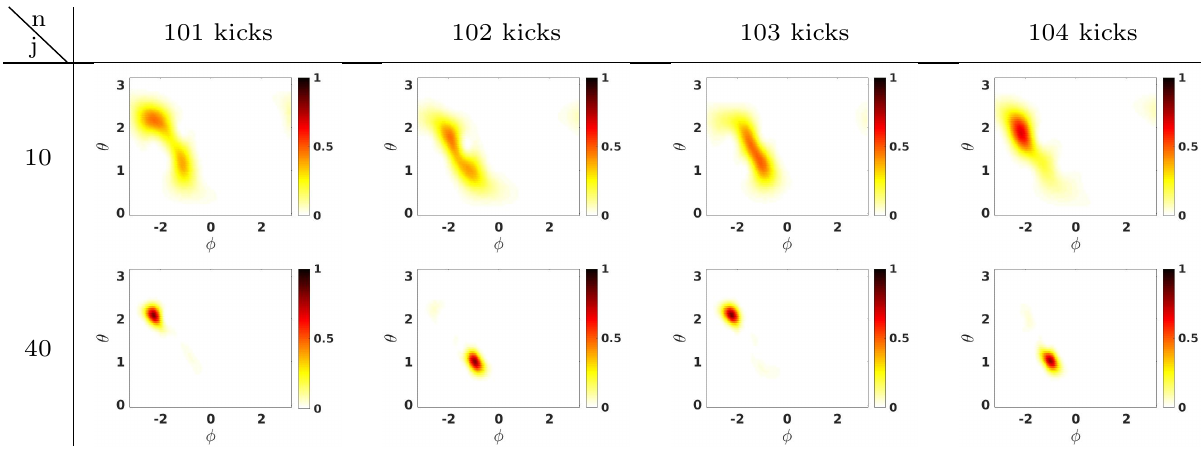}
    \caption{Evolution of the Husimi phase space distribution of an SCS centered on a point in $P2_A$ for two different $j$ values with $\kappa=2.5$. For $j=10$ (top row), the quantum-classical correspondence is weak compared to $j=40$.} 
    \label{k3}
\end{figure*}

In the range, $2 \leq \kappa< \pi$, we have two more fixed points, $FP_3$ and $FP_4$, and a period-2 orbit, $P2_A$, in addition to the ones present for $\kappa < 2$, as explained in Sec. \ref{sec:level2a}. $FP_1$ and $FP_2$ are unstable in this range while all others are stable. $FP_3$, $FP_4$ and $P2_A$ are functions of $\kappa$ (Table \ref{CP} ). For $\kappa=2.5$, the overlap between the spin coherent states centered on the two points in $P2_A$ is on the order of $10^{-4}$ for $j=10$, and on the order of $10^{-14}$ for $j=40$. Correspondingly, we see in Fig.~\ref{k3} that for $j=40$, the quantum dynamics follows the classical dynamics more closely, compared to $j=10$.  

\subsection{\label{sec:level5c}{Effect of classical instability}}

\begin{figure}
    \centering
    \includegraphics[scale=0.8]{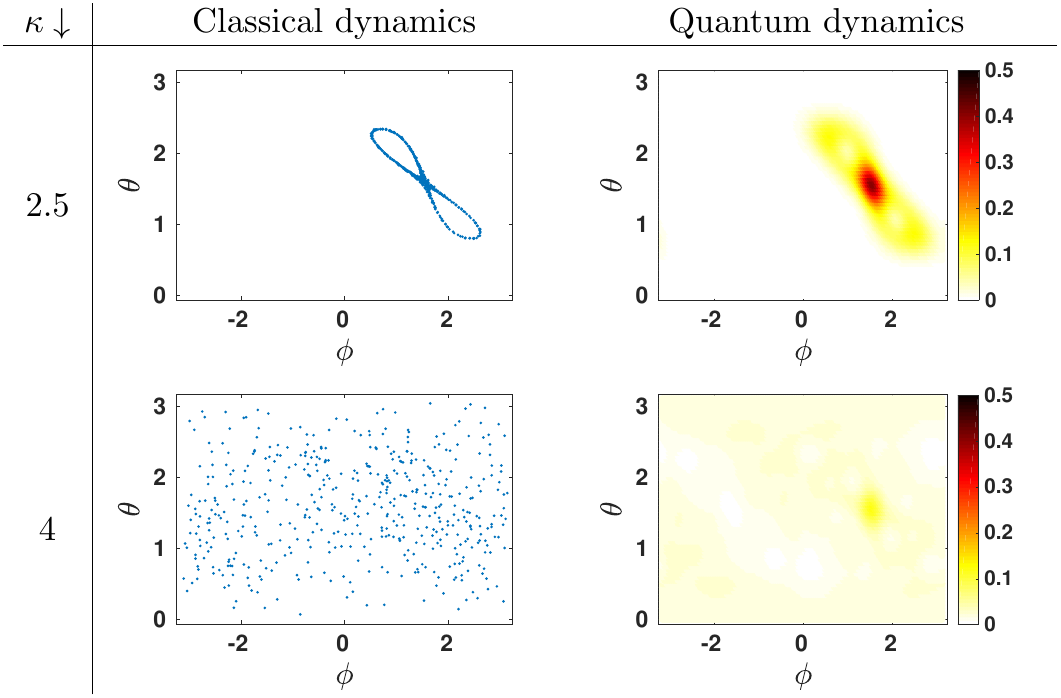}
    \caption{Effect of classical instability on quantum dynamics. Left column: Classical trajectories for an initial state slightly perturbed from the unstable fixed point, $FP_1$, in Table \ref{CP}, for two different $\kappa$. Right column: Husimi phase space distribution (averaged over 100 kicks) corresponding to an initial $j=25$ SCS centered on $FP_1$, for two different $\kappa$. }
    \label{k4}
\end{figure}

Classical instability of any periodic orbit leads to exponential divergence of classical trajectories even for infinitesimally small differences in initial conditions. We show in Fig.~\ref{k4} that when the quantification criteria is satisfied, the quantum initial states localized at an unstable fixed point explores the same regions of the Husimi phase space as  classical initial states slightly perturbed from an unstable fixed point would explore. This shows that classical instability affects classical as well as quantum dynamics in similar ways. We have checked this correspondence for the period-4 orbit, $P4$ in Table \ref{CP}, as well. The period-4 orbit is unstable for $\kappa>\pi$. The phase space is predominantly chaotic for $\kappa>3.5$ except for the regular islands of $FP_3, FP_4$ and $P2_A$. Thus, any classical initial state perturbed from the period-4 orbit explores most of the phase space avoiding the aforementioned regular islands. We have observed the same behavior for quantum initial states centered close to any point on the period-4 orbit for $\kappa>3.5$, that is, such a quantum state spreads out in the Husimi phase space avoiding the regions of classical regular islands.

\section{\label{sec:level6}Quantum signatures of classical bifurcations}

Given our new criteria for quantum-classical correspondence, we would ideally like to use it to identify classical bifurcation behavior (as shown in Fig. \ref{bifur_diagram}) in the quantum dynamics. To do so, we first define a measure of quantum dynamics that we can use to explore bifurcations. 
The survival probability of a quantum state, $|\psi(0) \rangle$, at time $t$, evolving according to a unitary operator, $U(t)$ is given by $|\langle \psi(0)|\psi(t) \rangle |^2$, where $|\psi(t)\rangle = U(t) |\psi(0) \rangle$. We analyze here the time-averaged survival probability of quantum states of the kicked top centered on any point of a classical period-$n$ orbit, where $n\geq 1$. 
\begin{enumerate}
\item Given a classical fixed point, we compute the quantity, 
\begin{equation}
S(L) = \frac{1}{L} \sum_{l=1}^L |\langle \psi(0)|\psi(l) \rangle |^2, 
\label{Period1}
\end{equation}
for some $L$, where $|\psi(l) \rangle =U^l |\psi(0) \rangle $, and $|\psi(0) \rangle$ is the SCS centered on the classical fixed point. Here, U is the unitary operator for one time period of the Floquet system.
\item Given any classical period-$n$ orbit, if $F$ denotes the classical map, then each of the $n$ points of the period-$n$ orbit will be a fixed point of the map, $F^n$. Thus, we study the survival probability of an SCS centered on any point of a classical period-$n$ orbit using the unitary operator, $U^n$, instead of $U$. For a classical period-$n$ orbit, we compute the quantity
\begin{equation}
S(L) = \frac{1}{L} \sum_{l=1}^L |\langle \psi(0)|\psi(nl) \rangle |^2, 
\label{HighPeriod}
\end{equation}
for some $L$, where $|\psi(nl) \rangle =U^{nl} |\psi(0) \rangle $, and $|\psi(0) \rangle$ is the SCS centered at any point of the classical period-$n$ orbit.
\end{enumerate}

\begin{figure}
\centering\includegraphics[scale=1.4]{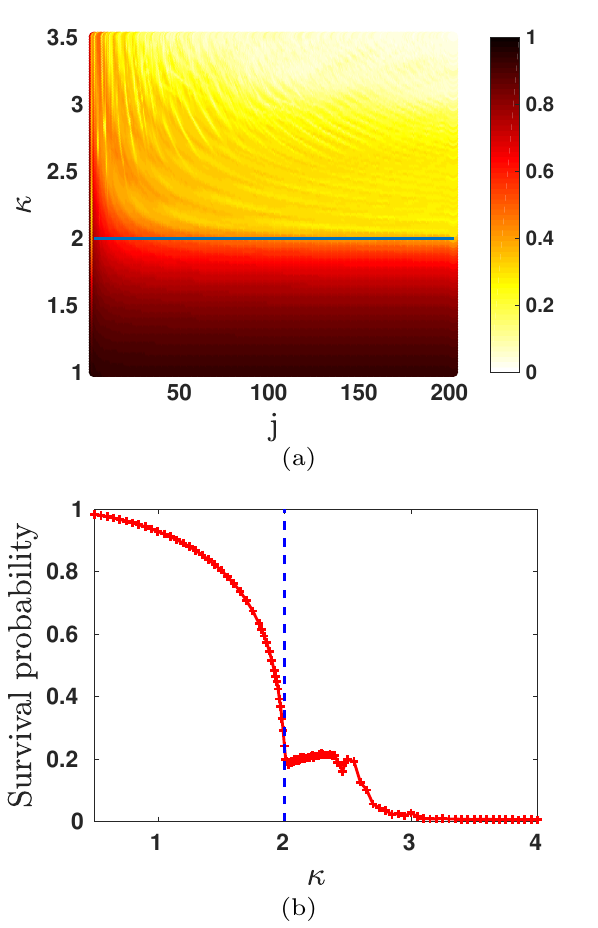}
\caption{(a) Survival probability of an SCS initially centered on $FP_1$ (in Table \ref{CP}), averaged over 50 kicks as a function of $j$ and $\kappa$. The horizontal line depicts the classical bifurcation. Darker color represents higher survival probability in this plot. (b) Survival probability of an SCS initially centered on $FP_1$, averaged over 200 kicks for $j=2000$ as a function of $\kappa$. The vertical dashed line represents the point of classical bifurcation.} 
\label{SurvFP1}
\end{figure}
\def \scalevalue {.58} 

\begin{figure}
\centering\includegraphics[scale=1.4]{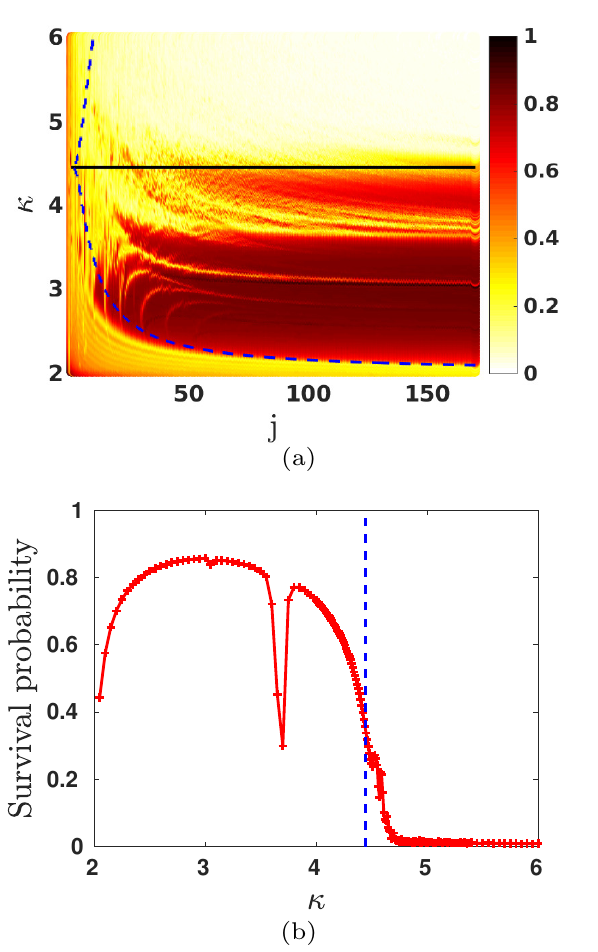}
\caption{(a) Survival probability of SCS initially centered on $P2_A$ (in Table \ref{CP}), averaged over 50 kicks as a function of $j$ and $\kappa$. The horizontal line depicts the classical bifurcation, and the dashed curve represents the $j$ value for a given $\kappa$ at which the overlap between the 2 SCS states corresponding to $P2_A$ is $\leq 10^{-10}$. Darker color represents higher survival probability in this plot. (b) Survival probability of SCS initially centered on $P2_A$ averaged over 100 kicks for $j=1000$ as a function of $\kappa$. The vertical dashed line represents the point of classical bifurcation.} 
\label{SurvP2A}
\end{figure}

\begin{figure}
\centering\includegraphics[scale=1.4]{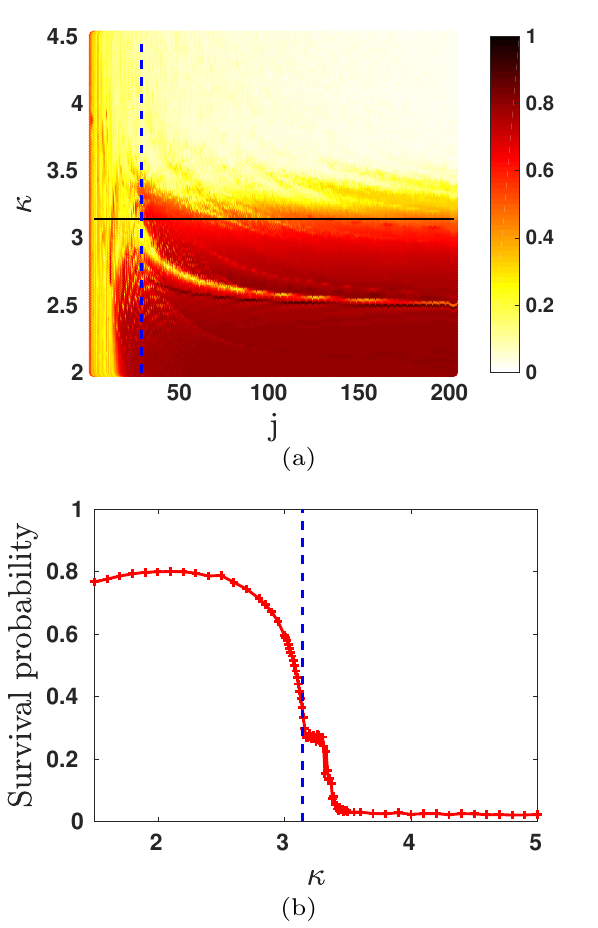}
\caption{(a) Survival probability [Eq.\ref{HighPeriod}] of SCS initially centered on $P4$ (in Table \ref{CP}), averaged over 50 kicks as a function of $j$ and $\kappa$. The horizontal line depicts the classical bifurcation, and the vertical dashed line represents the $j$ value at which the overlap between any two of the four SCS states corresponding to $P4$ is $\leq 10^{-8}$. Darker color represents higher survival probability in this plot. (b) Survival probability [Eq.\ref{HighPeriod}] of SCS initially centered on $P4$ averaged over 50 kicks for $j=1000$ as a function of $\kappa$. The vertical dashed line represents the point of classical bifurcation.} 
\label{SurvP4}
\end{figure}

We have plotted the survival probabilities corresponding to $FP_1$, $P2_A$ and $P4$ of Table \ref{CP} in Figs. \ref{SurvFP1}, \ref{SurvP2A} and \ref{SurvP4} respectively. Each figure consists of two plots, one illustrating signatures of bifurcation in the deep quantum regime, and the other in the semiclassical regime. 

(a) Analysis of $FP_1$ (Fig.~\ref{SurvFP1}): 
We see clear signatures of classical bifurcation of $FP_1$ at $\kappa=2$ in the survival probability plots in Fig.~\ref{SurvFP1} in a deep quantum regime as well as the semiclassical regime. The quantum state remains localized at the fixed point prior to bifurcation (because the fixed point is stable prior to bifurcation) and gets delocalized after bifurcation. 

(b) Analysis of $P2_A$ (Fig.~\ref{SurvP2A}): 
In Fig. \ref{SurvP2A}~(a) the classical bifurcation point (solid horizontal line) is easy to identify in the survival probability plot. Above this line, the survival probability is small, indicating that bifurcation has occurred. Below the horizontal line, however, there is some structure in the behavior of the survival probability. This can be understood in the following way. The two points associated with the period-2 orbit, $P2_A$ (Table \ref{CP}), are $\kappa$-dependent. Thus, the $j$ value at which the two SCS centered at these two points are orthogonal to each other is also $\kappa$-dependent. The dashed curve in Fig. \ref{SurvP2A}(a) represents the $j$ value at which the overlap between the two aforementioned SCS is less than $10^{-10}$ for the corresponding $\kappa$ values. Below this curve, we see small survival probability. This is because of mixing of the quantum dynamics between the two SCS states because they are not orthogonal to each other. Hence for $j$ values below the dashed curve the quantum dynamics does not mimic the corresponding classical dynamics in the period-2 orbit. Above the dashed curve, the quantum and classical dynamics should track, so there should be high survival probability (darker regions in the plot) below the classical bifurcation (solid line), and low survival probability above the bifurcation line. However there are also some lighter regions of low probability below the bifurcation line. One of the reasons for this is the quantum phenomenon of dynamical tunneling. Both the points of $P2_A$ are fixed points for the square of the classical map of the kicked top, thus allowing for dynamical tunneling between the two in addition to the period-2 motion between the two points. 

In Fig. \ref{SurvP2A}(b), for $j=1000$ (semiclassical regime), the bifurcation at $\kappa = \sqrt{2}\pi$ is clearly visible. The initial dip in the curve close to $\kappa = 2$ is because of non-zero overlap between the two aforementioned SCS states for $\kappa$ very close to 2. We also see a surprising dip in the survival probability around $\kappa=3.7$ though the $P2_A$ orbit is still stable. Further investigation of the classical phase space of the kicked top near this value of $\kappa$ reveals that a period-6 orbit arises near this period-2 orbit around $\kappa=3.62$. This period-6 orbit breaks off to the chaotic sea near $\kappa=3.68$, which results in the period-2 island in the phase space becoming smaller around $\kappa=3.68$. Thus, the wave packet centered at the period-2 orbit delocalizes to some extent in the phase space around $\kappa=3.68$. The size of the period-2 island increases again beyond $\kappa=3.72$ which results in a higher survival probability beyond $\kappa=3.72$ until bifurcation occurs. Dynamical tunneling also occurs around $\kappa=3.7$ to some extent, though the sum of the survival probability at the two points of the periodic orbit is not very close to 1 because of delocalization. These two points explain the dip at $\kappa=3.7$ in Fig. \ref{SurvP2A}(b). As $\kappa$ increases, we clearly observe very small survival probability after the classical bifurcation point in the deep quantum regime as well as the semiclassical regime.

(c) Analysis of $P4$ (Fig.~\ref{SurvP4}): The overlap between each pair of the four points associated with the period-4 orbit, $P4$ (Table \ref{CP}), is less than $10^{-8}$ for $j\geq 27$. As explained for $P2_A$, mixing of dynamics can happen in $P4$ for small $j$ values. For larger $j$ compared to the critical value of $j$, we observe a clear signature of bifurcation in the survival probability plots. 

We also note some general observations about the signatures of classical bifurcations in the quantum dynamics. The quantum dynamics changes smoothly with the classical bifurcations, unlike the classical dynamics which shows a sudden change. When any local bifurcation occurs that gives rise to new fixed points or periodic orbits, these new orbits are usually close to each other in the classical phase space, due to which the corresponding coherent states are not orthogonal to each other. As the bifurcation parameter (external control parameter) is varied, these new orbits get further apart which decreases the overlap between the corresponding coherent states. Eventually, they may become orthogonal at which point the correspondence between classical and quantum dynamics near these orbits is  restored (as long as other bifurcations do not occur prior to it). Alternatively, one could have increased the quantum number keeping the value of the external control parameter fixed. This explains why the quantum dynamics is affected smoothly by a classical bifurcation which gives rise to new fixed points or periodic orbits. 
Far from the classical bifurcation points, the stability of the classical fixed points and periodic orbits affects classical and quantum dynamics in the same way as long as the  quantification criteria given in Sec. \ref{sec:level4} are satisfied.
\section{\label{sec:level7}Quantum-classical correspondence in chaotic systems}
The Ehrenfest break time after which quantum and classical dynamics diverge is very small for chaotic systems, even in the semiclassical limit. Using our quantification criteria in Sec. \ref{sec:level5}, we explain the reason for short break times in classically chaotic systems whose quantum counterpart is finite-dimensional. There are various routes to classical chaos, such as the period-doubling route and the intermittency route \cite{hilborn2000chaos}. In the period-doubling route to chaos, there exists at least one $2^n$ periodic orbit for every $n$ at the onset of chaos. According to our quantification criteria, if there exists a periodic orbit with, say, $r$ periodicity in a classical system, then the quantum system needs to be at least $r$-dimensional to exhibit a correspondence with the classical dynamics in the vicinity of that periodic orbit. This is because the coherent states corresponding to all the points in any period-$n$ orbit need to be orthogonal to each other for correspondence between classical and quantum dynamics. Since in a period-doubling route to chaos, there exists periodic orbits with at least one $2^n$ periodic orbit for every $n$ at the onset of chaos, any finite-dimensional quantum system cannot exhibit correspondence with the classical dynamics even in the semiclassical limit because the dimension of the set of orthogonal quantum states in such a system is restricted by the dimension of the basis of the corresponding Hilbert space. Since there will always exist periodic orbits with periodicity higher than the dimension of the Hilbert space at the onset of chaos, there cannot exist a good correspondence between the classical and quantum dynamics at the onset of chaos given that the route to chaos generates periodic orbits of unbounded periodicity. This explains why we have a short break time for chaotic systems.
\section{\label{sec:level8}Conclusion}
Quantum-classical correspondence for chaotic systems and for systems with a mixed phase space has remained a long-standing open question. Periodic orbits, and their stability and bifurcations play an important role in the transition from regular to chaotic behavior. Thus, gaining insight into quantum-classical correspondence in the vicinity of periodic orbits, and understanding the role of stability of periodic orbits and bifurcations on the quantum-classical correspondence is of vital importance. We have proposed the conditions under which the coherent states, which are the most classical states in quantum, evolve in close conjunction with classical dynamics for Floquet systems. We have applied our criteria to the quantum kicked top and showed how it can be used to quantify Bohr's correspondence principle. We note that in some situations quantum and classical dynamics may correspond even if our conditions are not met, but in general this will not be the case. Our studies of the kicked top seemed to indicate that such exceptions are not common. We have also illustrated the effect of classical instability on quantum evolution. Our analysis shows that the survival probability of quantum states centered on the periodic orbits exhibits signatures of classical bifurcations, given the aforementioned criteria are satisfied. Furthermore, we have used our criteria for quantum-classical correspondence to explain the reason for short break times between quantum and classical dynamics in chaotic systems. 

Signatures of chaos have been widely studied in the quantum kicked top using various quantum theoretic measures in the deep quantum regime as well as the semiclassical regime \cite{Fox1994,entanglement2004PRE,entanglement2004PRA, Chaos2008,lombardi2011,discord2015,santhanam2017,pattanayak2017}. Our analysis and criteria for quantum-classical correspondence can be applied to understand these previous results and develop better quantum control techniques. Furthermore, our criteria can be experimentally tested using current technology.

\begin{acknowledgments}
M. K. and S.G. acknowledge support from the Discovery Programme of the National Science and Engineering Research Council of Canada (NSERC).
\end{acknowledgments}

\end{document}